\pdfoutput=1

\documentclass[a4paper, 11pt]{article}

\usepackage{hyperref}
\usepackage{latexsym, graphicx}
\usepackage{amsmath, amssymb, bm, longtable, array, tabularx, subcaption}
\usepackage{appendix}
\usepackage{verbatim}
\usepackage{cite}

\usepackage{xcolor}

\renewcommand\tilde{\widetilde}
\newcommand{\vphi}{ \bm{\phi} }

\begin{document}
\thispagestyle{empty}
\renewcommand{\thefootnote}{\fnsymbol{footnote}}

\begin{titlepage}

\begin{flushright}
USTC-ICTS-15-12
\end{flushright}

\vspace{0.2in}

\begin{center}

\textbf{
{\Large Flatness of Minima in Random Inflationary Landscapes}
}

\vspace{0.2in}

\textbf{
Yang-Hui He$^{a,b,c,}$\footnote{\texttt{hey@maths.ox.ac.uk}},
Vishnu Jejjala$^{d,}$\footnote{\texttt{vishnu.jejjala@gmail.com}},
Luca Pontiggia$^{d,}$\footnote{\texttt{lucatpontiggia@gmail.com}},
Yan Xiao$^{a,}$\footnote{\texttt{Yan.Xiao@city.ac.uk}},
Da Zhou$^{a,e,}$\footnote{\texttt{zhouda@mail.ustc.edu.cn}}\\
}

\vspace{0.1in}

${}^a$ \textit{Department of Mathematics, City, University of London,\\
Northampton Square, London EC1V 0HB, UK}
\vskip0.25cm

${}^b$ \textit{Merton College, University of Oxford, OX1 4JD, UK}
\vskip0.25cm

${}^c$ \textit{School of Physics, NanKai University, Tianjin, 300071, P.R.\ China,}
\vskip0.25cm

${}^d$ \textit{Mandelstam Institute for Theoretical Physics, NITheP, CoE-MaSS,} and \textit{School of Physics,\\
University of the Witwatersrand, Johannesburg, WITS 2050, South Africa}
\vskip0.25cm

${}^e$ \textit{The Interdisciplinary Center for Theoretical Study,\\
University of Science and Technology of China, Hefei, Anhui, 230026, China}

\end{center}

\vspace{10mm}

\begin{abstract}
\noindent We study the likelihood for relative minima of random polynomial potentials to support the slow-roll conditions for inflation.
Consistent with renormalizability and boundedness, the coefficients that appear in the potential are chosen to be order one with respect to the energy scale at which inflation transpires.
Investigation of the single field case illustrates a window in which the potentials satisfy the slow-roll conditions.
When there are two scalar fields, we find that the probability depends mildly on the choice of distribution for the coefficients.
A uniform distribution yields a $0.05\%$ probability of finding a suitable minimum in the random potential whereas
a maximum entropy distribution yields a $0.1\%$ probability.
\end{abstract}

\end{titlepage}

\tableofcontents

\newpage

\renewcommand{\thefootnote}{\arabic{footnote}}
\setcounter{footnote}{0}

\section{Introduction}

In order to solve the well known horizon and flatness problems, cosmological inflation~\cite{Cosmo_Chap1,Cosmo_Intro5,Cosmo_Intro6,Cosmo_Intro7} posits that the Universe underwent a period of exponential expansion early in its history.
To date, there is no uniquely compelling realization of how inflation transpired.
The literature abounds with numerous and varied proposed mechanisms~\cite{Cosmo_Chap2, Cosmo_Intro8}.
Paradigmatic models involve scalar fields which dynamically roll until arriving at the (relative) minimum of some potential.

While a model of physics that purports to approximate our world must correctly trace out the cosmological history of the Universe, these are not the only considerations in selecting a theory.
The Standard Model of particle physics establishes that at low energies the particles in Nature organize themselves into three generations of chiral fields that transform in representations of the $SU(3)\times SU(2)_L\times U(1)_Y$ gauge group.
Top down realizations of low energy gauge theories from a fundamental theory such as string theory typically augment the symmetries of the S-matrix with supersymmetry.
The simplest scenario for preserving ${\cal N}=1$ supersymmetry in four dimensions involves the compactification of the heterotic string in ten dimensions on a Calabi--Yau threefold~\cite{CY_Chap1}.
This effort has led to a number of constructions that reproduce the matter spectrum and Yukawa interactions observed in the Standard Model~\cite{CY_Chap23,Cosmo_Chap3,CY_Chap24,Cosmo_Chap4,Cosmo_Chap5}.
Again, we have an abundance of models that are \textit{a priori} indistinguishable on the basis of experiments.

As we do not have a \textit{sui generis} path to the real world, we propose to study a large class of models at once and incorporate inputs of both cosmology and particle physics.
The most important characterization of a Calabi--Yau threefold is a pair of topological invariants $h^{1,1}$ and $h^{2,1}$.
There are $h^{1,1}$ K\"ahler and $h^{2,1}$ complex structure parameters that describe the size and the shape of the geometry.
In the most na\"{\i}ve setup, these deformation parameters supply candidates for the scalar fields in inflation.
The largest available catalog of Calabi--Yau threefolds is derived from the Kreuzer--Skarke database of reflexive polytopes~\cite{CY_Chap8,ksd}.
Using the methods of Batyrev and Borisov~\cite{CY_Chap6,CY_Chap7}, each consistent triangulation of a reflexive polytope yields a toric Calabi--Yau manifold.
In~\cite{CY_Chap16,Cosmo_Chap7}, topological and geometric data are tabulated for the Calabi--Yau threefolds thus obtained for low values of $h^{1,1}$.
Heterotic Standard Model constructions in string theory typically employ Calabi--Yau geometries with small values of the Hodge numbers.
For example,~\cite{Cosmo_Chap3} uses a manifold with $(h^{1,1},h^{2,1})=(3,3)$.
Where there are explicit candidates for particle physics from string theory, we expect only a small number of moduli to appear in the low energy effective action.
Motivated by this fact, these are the models that we investigate in this article.

We aim to provide statistics for how many (possibly metastable) vacua support slow-roll constraints on inflation.
Working in effective field theory, we examine random polynomial potentials for inflation with a small number of scalar fields.
The justification for examining these models derives from string constructions of de Sitter like metastable vacua, \textit{e.g.}, the KKLT~\cite{Cosmo_Chap8} and Large Volume Scenarios~\cite{Cosmo_Chap9}.
In the latter class of models, the number of flat directions in the low energy effective potential is given by the number of parametrically large four cycles.
(In fact, the number of flat directions is one less than the number of large cycles~\cite{Cosmo_Chap9}.)
There are $69$ explicit Calabi--Yau geometries with two large cycles and one known Calabi--Yau geometry with three large cycles~\cite{Cosmo_Chap10}.
These are candidate manifolds for bona fide cosmological model building in string theory that correspond to one field and two field inflation.
In light of this, we study random potentials relevant to these cases in particular.

The potentials we study are sums of monomials in the scalar fields.
We truncate the expansion to focus on the interactions that are relevant or marginal from the perspective of a four dimensional low energy effective action.
As higher order monomials in the fields are irrelevant operators, we expect these to be mass suppressed and neglect them for the purposes of our investigations.
We will study models where the time scale for inflation if $t_\mathrm{GUT}$.
Correspondingly, the energy scale in the problem is $M_\mathrm{GUT}$. 
Invoking naturalness~\cite{Cosmo_Chap11}, we choose coefficients in the potential to be order one with respect to this scale.
(It is not always required that we choose $t_\mathrm{GUT}$ as the natural time scale in the problem; in fact, certain models~\cite{Cosmo_Chap2, Cosmo_Intro8} employ energy scales which are lower than $M_\mathrm{GUT}$.)
In discussing single field models with order one coefficients, our work is analytic.
In multi-field models, we construct random potentials whose coefficients are selected from distributions.

At the outset, we should note that the cases we analyze where there are few moduli may well represent an atypical class of string compactifications.
While there are $473\,800\,776$ reflexive polyhedra in four dimensions, there are only $30\,108$ pairs of Hodge numbers that appear in the threefold dataset.
The number of reflexive polytopes in the Kreuzer--Skarke list peaks at the Hodge numbers $(h^{1,1},h^{2,1}) = (27,27)$.
There are $910\,113$ such polytopes.
Indeed, there are significant and surprising patterns in the distribution of Calabi--Yau geometries close to this maximum~\cite{Cosmo_Chap12}.
Reflexive polytopes with low Hodge numbers are sparse in the Kreuzer--Skarke database.
Flux compactifications on Calabi--Yau threefolds yield, in principle, an enormously large number of potential vacua for string theory~\cite{Cosmo_Chap13,Cosmo_Chap14}.
There are, however, to date no explicit constructions of the Standard Model on a geometry with Hodge numbers that correspond to those of a typical Calabi--Yau manifold.
As there are good reasons to be skeptical about anthropic resolutions to the cosmological constant problem and there are potential issues regarding the stability of the flux vacua~\cite{Cosmo_Chap15,Cosmo_Chap16,Cosmo_Chap17}, we adopt an agnostic attitude.
We simply note that if a construction is stable in this context, a generic compactification on a typical Calabi--Yau manifold will most likely involve a large number of moduli fields.
As we review below, the large-$N$ limit of inflaton fields is studied in complementary work.

The organization of the paper is as follows.
In Section~\ref{sec:random}, we discuss the setup for random inflation.
In Section~\ref{sec:single}, we investigate the case of single field inflation with $O(1)$ coefficients.
This analysis is a completely analytic study of polynomial equations.
In Section~\ref{sec:multi-field}, we examine the case of two scalar fields with couplings up to quartic order.
Again, the coefficients are $O(1)$.
We choose coefficients using a uniform distribution and a Gaussian distributions for coefficients of indefinite sign and a gamma distribution for coefficients that are positive.
In Section~\ref{sec:discussion}, we remark on future investigations in the context of semi-realistic string models.

\section{Random potentials for inflation}\label{sec:random}

The action we consider takes the form
\begin{eqnarray}
  I = - \int d^4x \sqrt{-g}\ \left( \frac{R}{16\pi G} + \frac{1}{2} g^{\mu\nu} \partial_\mu \vphi \cdot \partial_\nu \vphi - V(\vphi) \right) ~,
  \label{eq:action}
\end{eqnarray}
where the Einstein--Hilbert term is supplemented by a matter sector that consists of $k$ scalar fields,
\begin{equation}
\vphi = (\phi^1, \phi^2, \cdots, \phi^k) ~.
\end{equation}
For simplicity, we assume that the metric in field space is the identity matrix, \textit{i.e.},
\begin{equation}
  \partial_\mu \vphi \cdot \partial_\nu \vphi = \delta_{ij} \partial_\mu \phi^i \partial_\nu \phi^j ~.
\end{equation}
The scalar potential $V(\vphi)$ determines the model and can be expanded as a polynomial in the fields $\phi^i$.

This setup lets us examine cosmological inflation.
Deducing the form of the potential is a long standing problem;
many scenarios present attractive phenomenological features, and to date observation has provided only limited guidance in selecting $V(\vphi)$.
In models with a single inflation field, famously the WMAP~\cite{Cosmo_Chap18} and Planck~\cite{Cosmo_Chap19} observations disfavor the simplest quadratic potential.
Other scenarios are variously consistent with the data.
See, for example,~\cite{Cosmo_Intro25} for a recent review.

One of the simplest multi-field models, hybrid inflation~\cite{Cosmo_Intro23}, involves coupling two fields according to the potential
\begin{align}
V(\phi,\psi) = \frac{1}{4\lambda}(\lambda\psi^2 - M^2)^2 +\frac{1}{2}m^2\phi^2 + \frac{\lambda'}{2}\phi^2\psi^2 ~.
\end{align}
Here, $\lambda$ and $\lambda'$ are couplings and $M$ and $m$ are the masses of $\psi$ and $\phi$, respectively.
We require that $V(\phi)=\frac{1}{2}m^2\phi^2 \ll \frac{M^4}{4\lambda}$.
This guarantees that the inflationary energy density of the false vacuum associated to the symmetry breaking potential $V(\psi)=\frac{\lambda}{4}(\psi^2 - M^2)^2$ dominates.
The effective mass for the $\psi$ field is $M_\mathrm{eff}^2 = -M^2 + \lambda' \phi^2$, which vanishes at $\phi_*^2 = M^2/\lambda'$.
Starting from $\phi^2\gg M^2$, the minimum is at $\psi = 0$.
This is morally a single field model with an effective potential of the form
\begin{align}
V_{\text{eff}} = \frac{\lambda}{4}M^4 +\frac{1}{2}m^2\phi^2 ~.
\end{align}
The field rolls until it reaches $\phi_*$.
The $\psi = 0$ locus is then unstable, and the field rolls again into the true minima at $\phi = 0$, $\psi = \pm M$.
Interest in the model stems from its versatility and success in predicting certain features of inflation, such as the power law behavior of the perturbation spectrum.
By tweaking the model in various ways, one can deal with inflation with or without first order phase transitions.
While this is a prototype multi-field model, there is a built-in hierarchy to the coefficients.
(See also~\cite{Cosmo_Chap20}.)

By constraining the inflationary scenario at a level matching the accuracy of current experimental data,~\cite{ Cosmo_Intro8} presents an encyclop{\ae}dia of $74$ satisfactory models.
In our work, we adopt a slightly different approach and address the question of how generic or specific the models should be in order to satisfy experimental constraints.
For this purpose, we consider randomly generated multi-field models (with the inflationary potential being given by polynomials with random coefficients) and verify whether the models can satisfy a certain set of conditions.
In particular, we demand that the scalar potential has a parameter window such that slow-roll conditions are satisfied.

Suppose there are some minima that satisfy the slow-roll conditions.
What are the global features of the potentials that accommodate this?
Turning the question around, given a large set of potentials (which may have some distribution in the function space), how likely is it that the potential has regions that satisfy slow-roll conditions?
How often can slow-roll inflation be accommodated with $O(1)$ coefficients?
These are the issues we aim to address below.

In an analysis of multi-field inflation, the need to establish the behavior of random potentials is almost compulsory.
Generic compacticiations, can have hundreds of scalar fields~\cite{CY_Chap4,Cosmo_Chap21,Cosmo_Chap22,Cosmo_Chap23}.
Since these theories describe physics at energy scales close to the inflationary scale, there is considerable interest in analyzing their dynamics.
Considering random potentials with large-$N$ fields has a considerable history~\cite{Cosmo_Intro20,Cosmo_Chap24,Cosmo_Chap25,Cosmo_Intro12,Cosmo_Intro21,Cosmo_Intro22,Cosmo_Chap26,Cosmo_Chap27,Cosmo_Chap28,Cosmo_Intro4,Cosmo_Chap29,Cosmo_Chap30,Cosmo_Chap31}.
As multi-field models have an almost infinite number of ways to inflate, the task of understanding how the potential energy driving inflation is distributed among all these fields becomes an incredibly difficult one.
This problem is often referred to as the measure problem, and it deals with attempting to handle the possible initial conditions~\cite{Cosmo_Intro10,Cosmo_Intro13,Cosmo_Intro9,Cosmo_Intro11}.
Multidimensional landscapes may also be afflicted by instabilities~\cite{Cosmo_Chap32,Cosmo_Chap33,Cosmo_Chap34}.
In general, the approach that random multi-field inflation adopts, is to study the dynamics of inflation by creating an ensemble of random potentials.
Then, through a statistical analysis, one can comment on the inflationary landscape produced by the respective models.
Related studies have recently appeared in the context of Gaussian models~\cite{Cosmo_Chap35,Cosmo_Chap36} and non-minimal kinetic terms~\cite{Cosmo_Chap37}.\footnote{
As we were completing this work, a similarly themed investigation appeared in~\cite{Cosmo_Chap38}.
This work examines inflationary landscapes corresponding to one dimensional potentials.}
The study of random potentials is of course not limited to inflation.
It is useful to borrow techniques for the generation of random potentials from other fields in physics, in particular string theory and quantum field theory~\cite{Cosmo_Chap39,Cosmo_Chap40}, and adapt these ideas to the cosmological context.

In Section~\ref{sec:single}, we analyze the single field case analytically.
In Section~\ref{sec:multi-field}, we investigate the statistics of random inflation by examining a large set of sample potentials for two field inflation.
In each potential, the coefficients are random numbers that fall within a particular range.
For each sample potential, we examine whether it has slow-roll regions.
We calculate the fraction of potentials that do have slow-roll regions and examine what features they have in common.
For succinctness, in the following we will use the term ``\emph{slow-roll potentials}'' to refer to those potentials that satisfy the slow-roll conditions in some region of the field space.
We assume the potential term $V(\vphi)$ is a polynomial in $\phi^i$ up to degree four and is bounded below.
In this paper, we only consider single field and two field inflation models.
In the former case, we shall denote $\vphi=\tilde\varphi$, and in the latter, $\vphi=(\tilde\varphi,\tilde\psi)$.
Both cases are developed in the following sections.

\section{Single field models}\label{sec:single}

The polynomial potential up to degree four for single field inflationary models has the form
\begin{eqnarray}
  V_{a,b}(\tilde\varphi) = \frac{\tilde\varphi^2}{2} (M^2 - a M \tilde\varphi + b \tilde\varphi^2) ~,
  \label{eq:single-V}
\end{eqnarray}
where $M\lesssim 10^{16}$ GeV is the mass of the inflaton $\tilde\varphi$, and $a$ and $b$ are two dimensionless random numbers. Note that in order for the potential to be bounded from below, the quartic term must be positive, which means we presume $b$ is positive.
Another feature about this potential is the symmetry
\begin{eqnarray}
  V_{-a,b}(\tilde\varphi) = V_{a,b}(-\tilde\varphi) ~,
\end{eqnarray}
which indicates if $V_{a,b}$ is a slow-roll potential, $V_{-a,b}$ must also be slow-roll.
So again we only need to assume that $a$ is positive.

Now, to factor out the parameter $M$, we perform a rescaling, $\tilde\varphi=M\varphi$, which also makes $\varphi$ dimensionless.
Then the potential can be recast as
\begin{eqnarray}
  V(\varphi) = \frac{M^4}{2} \varphi^2 (1 - a\varphi + b\varphi^2) ~,
  \label{eq:new-single-V}
\end{eqnarray}
where we have omitted the two subscripts $a$ and $b$ on $V$.\footnote{
Adding a zeroth order term to the potential will shift the energy of the relative minimum. Though the potential appears in the denominator of the slow-roll conditions, we assume constant terms in the potential do not greatly affect flatness and therefore neglect such a term in writing the potential.}
Consequently, the two slow-roll parameters are
\begin{eqnarray}
  \epsilon &=& \frac{M_\mathrm{Pl}^2}{2} \left( \frac{V'(\tilde\varphi)}{V(\tilde\varphi)} \right)^2
  = \frac{1}{2\mu} \left( \frac{V'(\varphi)}{V(\varphi)} \right)^2 ~,\nonumber\\
  \eta &=& M_\mathrm{Pl}^2 \frac{V''(\tilde\varphi)}{V(\tilde\varphi)}
  = \frac{1}{\mu} \frac{V''(\varphi)}{V(\varphi)} ~,
  \label{eq:slow-roll-parameters}
\end{eqnarray}
where $M_\mathrm{Pl}$ is the Planck mass and $\mu=M^2/M_\mathrm{Pl}^2$ is the square of the ratio between the mass of the inflaton and the Planck mass.
The Planck 2015~\cite{Cosmo_Intro25} data tells us that the scalar spectral index is measured to be $n_s = 0.9655\pm 0.0062$ and the slow-roll parameters are deduced to satisfy
\begin{eqnarray}
\epsilon < 0.012 ~, \qquad \eta = -0.0080^{+0.0088}_{-0.0146} ~.
\end{eqnarray}
Noting that $n_s - 1 = 2\eta - 6\epsilon$, in our analysis we demand that the slow-roll parameters are $O(10^{-2})$:
\begin{eqnarray}
  \epsilon < 0.01 ~,\qquad |\eta| < 0.01 ~.
  \label{eq:slow-roll-condition}
\end{eqnarray}
Given the definition of $\mu$, the two slow-roll
parameters are actually independent of the specific value of inflaton mass $M$ because the
$M^4$ term in~\eqref{eq:new-single-V} appears in both the numerator and denominator of~\eqref{eq:slow-roll-parameters} and therefore cancels.

If we define a new variable $y=a\varphi$, the whole analysis will only depend on the ratio
of $b$ to $a^2$, which shall be dubbed $\beta$, instead of the explicit values of $a$
and $b$. So we can define an auxiliary potential,
\begin{eqnarray}
  v(y) = \frac{2a^2}{M^4} V(\varphi) = y^2 (1 - y + \beta y^2) ~, \qquad \beta = \frac{b}{a^2} > 0 ~,
  \label{eq:single-auxiliary-v}
\end{eqnarray}
and two new slow-roll parameters which only depend on one parameter $\beta$,
\begin{eqnarray}
  \bar\epsilon = \frac{1}{2} \left( \frac{v'}{v} \right)^2 = \frac{\nu}{0.01} \epsilon ~,\qquad
  \bar\eta = \frac{v''}{v} = \frac{\nu}{0.01} \eta ~,
  \label{eq:new-slow-roll-param}
\end{eqnarray}
where
\begin{eqnarray}
  v' \equiv \frac{dv}{dy} ,\qquad v'' \equiv \frac{d^2v}{dy^2} ~,\qquad
  \nu \equiv \frac{0.01\mu}{a^2} = \frac{0.01}{a^2} \frac{M^2}{M_\mathrm{Pl}^2} ~.
  \label{eq:define-nu}
\end{eqnarray}
The slow-roll conditions become
\begin{eqnarray}
  \bar\epsilon < \nu ~,\qquad |\bar\eta| < \nu ~.
  \label{eq:new-slow-roll-cond}
\end{eqnarray}
Of course, we have assumed $a\neq 0$ in~\eqref{eq:single-auxiliary-v} and~\eqref{eq:new-slow-roll-cond},
and the special $a=0$ case can be approximated by setting $a$ to be an extremely
small nonzero number, then $a\to 0$ corresponds to the $\beta\to\infty$ case, which is
a special situation that will be discussed in Section~\ref{sec:shape-1}.

From~\eqref{eq:single-auxiliary-v} one can see that as $y\to\pm\infty$, $v\sim y^4$
while $v'\sim y^3$ and $v''\sim y^2$, so~\eqref{eq:slow-roll-condition} is always
satisfied. That means, there exists a $y_0>0$ and a $y'_0<0$ such that~\eqref{eq:slow-roll-condition} holds true for all $y>y_0$ or all $y<y'_0$.
In other words, in any cases there are always at least two trivial slow-roll regions,
$(-\infty,y'_0)$ and $(y_0,\infty)$.
Our search for an inflationary scenario excludes
these regions where the~\eqref{eq:slow-roll-condition} is satisfied simply due to the largeness of the potential
$v$.\footnote{
This is not to say that regions where inflation transpires by virtue of a large denominator $v$ must always be disregarded.
 Models such as chaotic inflation can use these trivial regions of the potential.}
 We aim to isolate other, perhaps more realistic scenarios which satisfy the slow-roll conditions with a flat $v$ (\textit{i.e.}, small $v'$ and $v''$) region of finite length. 

From \eqref{eq:new-slow-roll-param} we have
\begin{eqnarray}
  \frac{d\bar\epsilon}{dy} = \frac{v'}{v^3} (v'' v - v'^2) =
  - \frac{y^2}{v^3} v' \big[ 2 - 4y + (3+2\beta) y^2 - 6by^3 + 4\beta^2 y^4 \big] .
  \label{eq:d-epsilon}
\end{eqnarray}
For $y<0$, we have $v>0$, $v'<0$, and the expression within the brackets is positive,
so $\frac{d\bar\epsilon}{dy}>0$. This means, if we find a $y'_0<0$ such that
$\bar\epsilon(y'_0)=0.01$, then $(-\infty,y'_0)$ is a trivial slow-roll region and
$(y'_0,0)$ does not contain a non-trivial slow-roll region.
Hereafter, we are only interested in the region with $y>0$.

In deducing the regions that satisfy the slow-roll conditions in single field inflation, we do not need to perform a Monte Carlo analysis or scan over potentials with random coefficients.
It suffices to analytically examine a system of polynomial equations.
We look for the true minimum of the potential.
Complementary investigations (see, for example,~\cite{Cosmo_Chap38}) examines relative minima in random landscapes.
In the following subsection, we will see that different intervals for $\beta$ exhibit characteristic behavior.

\subsection{Behavior of slow-roll parameters}

Graphically, we can draw the slow-roll parameters, which are functions of $y$ given a
specific value of $\beta$, on the plane and use the horizontal lines $\bar\epsilon=\nu$
and $\bar\eta=\pm\nu$ to intercept curves of the slow-roll parameters $\bar\epsilon$ and
$\bar\eta$ respectively, then from the interception one can easily read off whether
there are slow-roll regions for the corresponding potential. The classification of different
behaviors of $\bar\epsilon$ and $\bar\eta$ will be represented below.

To determine the behavior of slow-roll parameters $\bar\epsilon$ and $\bar\eta$, we
compute their partial derivatives with respect to $y$,
\begin{eqnarray}
  \frac{\partial\bar\epsilon}{\partial y} = \frac{y^2 v'}{v^3} f(y, \beta) \ ,\qquad
  \frac{\partial\bar\eta}{\partial y} = \frac{2y}{v^2} g(y, \beta) ,
  \label{eq:d-eps-eta}
\end{eqnarray}
where
\begin{eqnarray}
  v' &:=& 2y - 3y^2 +4\beta y^3 ,\nonumber\\
  f(y,\beta) &:=& -2 + 4y - (3+2\beta) y^2 + 6\beta y^3 - 4\beta^2 y^4 ,\nonumber\\
  g(y,\beta) &:=& -2 + 6y - (6+4\beta) y^2 + 15\beta y^3 - 12\beta^2 y^4 .
  \label{eq:def-f-g}
\end{eqnarray}

\begin{figure}[!ht]
  \centering
  \includegraphics[width=.45\textwidth]{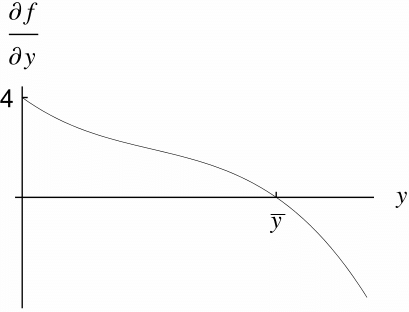} \quad
  \includegraphics[width=.45\textwidth]{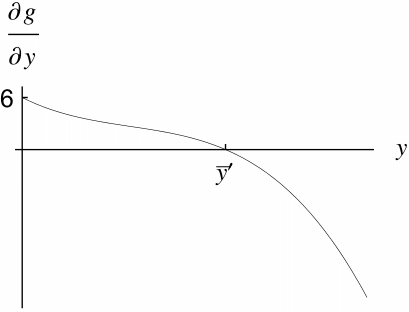}
  \caption{\textit{Roots of $\frac{\partial f}{\partial y}$ and $\frac{\partial g}{\partial y}$.}}
  \label{fig:df-dg}
\end{figure}

For $\beta>9/32$ we have $v>0$ and $v'>0$, so the signature of $\frac{\partial\bar\epsilon}{\partial y}$
or $\frac{\partial\bar\eta}{\partial y}$ depends on the signature of $f(y,\beta)$ or $g(y,\beta)$.
In order to determine the signature of $f$ and $g$, we differentiate them with respect to $y$,
\begin{eqnarray}
  \frac{\partial f}{\partial y} &=& 4 - 2(3+2\beta)y + 18\beta y^2 -  16\beta^2 y^3 ,\nonumber\\
  \frac{\partial^2 f}{\partial y^2} &=& -48(\beta y - 3/8)^2 - (4\beta - 3/4) < 0 ,\nonumber\\
  \frac{\partial g}{\partial y} &=& 6 - 2(6+4\beta)y + 45\beta y^2 - 48\beta^2 y^3 ,\nonumber\\
  \frac{\partial^2 g}{\partial y^2} &=& -144 (\beta y -5/6)^2 - (8\beta - 33/16) < 0 .
  \label{eq:d-f-g}
\end{eqnarray}
Since $\frac{\partial^2 f}{\partial y^2}<0$ and $\frac{\partial^2 g}{\partial y^2}<0$,
$\frac{\partial f}{\partial y}$ and $\frac{\partial g}{\partial y}$ are monotonically decreasing
functions of $y$. In addition, $\frac{\partial f}{\partial y}(y=0)>0$ and
$\frac{\partial f}{\partial y}(y=\infty)<0$, so $\frac{\partial f}{\partial y}$ has one root in
$(0,\infty)$. By the same token, $\frac{\partial g}{\partial y}$ also has one root in $(0,\infty)$.
We illustrate the previous analysis in Figure~\ref{fig:df-dg}.
From Figure~\ref{fig:df-dg} we can see that both functions $f$ and $g$ have one and only one
maximum which is $\bar{y}$ (respectively $\bar{y}'$) in $(0,\infty)$. As $f(0,\beta)=g(0,\beta)=-2$
and $f(\infty,\beta)$ or $g(\infty,\beta)<0$, we conclude, (1) if $f(\bar{y},\beta)<0$ (respectively,
$g(\bar{y}',\beta)<0$), $\bar\epsilon$ (respectively, $\bar\eta$) is monotonically decreasing in $(0,\infty)$;
(2) if $f(\bar{y},\beta)>0$ (respectively, $g(\bar{y}',\beta)>0$), $\bar\epsilon$ (respectively, $\bar\eta$)
has one local minimum and one local maximum in $(0,\infty)$. Between these two cases, there is
an intermediate stage which is $f(\bar{y},\beta)=0$ or $g(\bar{y}',\beta)=0$. As a result, we
need to solve following two sets of equations,
\begin{eqnarray}
  \left\{ 
    \begin{array}[]{rcl}
      f(y,\beta) & = & 0 \\
      \frac{\partial}{\partial y} f(y,\beta) & = & 0
    \end{array}
  \right.
  \qquad \textrm{and} \qquad
  \left\{ 
    \begin{array}[]{rcl}
      g(y,\beta) & = & 0 \\
      \frac{\partial}{\partial y} g(y,\beta) & = & 0
    \end{array}
  \right. .
  \label{eq:f-df-g-dg}
\end{eqnarray}
The two sets of equations are reduced, by a Groebner basis elimination, to
\begin{eqnarray}
  (784\beta^3 - 846\beta^2 + 270\beta - 27) (4\beta - 1) = 0
  \label{eq:f-df-reduce}
\end{eqnarray}
and
\begin{eqnarray}
  (8192\beta^3 - 10368\beta^2 + 3591\beta - 378) (25\beta - 6) = 0
  \label{eq:g-dg-reduce}
\end{eqnarray}
respectively. Eq.~\eqref{eq:g-dg-reduce} gives $\beta=0.778890$, which supplies the bounds for the interval in Section~\ref{sec:shape-1}, and~\eqref{eq:f-df-reduce} gives $\beta=0.602103$, which then supplies the bounds for the interval in Section~\ref{sec:shape-2}.

\subsubsection{$\beta\geq 0.7789$}
\label{sec:shape-1}

\begin{figure}[!htp]
  \centering
  \includegraphics[width=.55\textwidth]{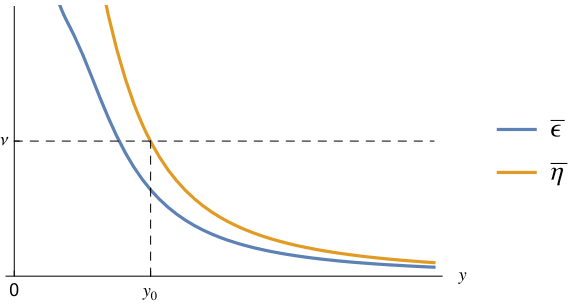}
  \caption{\textit{Shapes of $\bar\epsilon$ and $\bar\eta$ for $\beta\geq 0.7789$.}}
  \label{fig:shape-1}
\end{figure}

For $\beta\geq 0.7789$, both $\bar\epsilon$ and $\bar\eta$ are monotonically decreasing for
$y\in(0,\infty)$, which is shown in Figure~\ref{fig:shape-1}.
From this graph it can be readily seen that given any $\nu$ there is only one trivial slow-roll
region for $y>0$ which in this graph is $(y_0,\infty)$.

\subsubsection{$0.6021\leq\beta<0.7789$}
\label{sec:shape-2}

\begin{figure}[!htp]
  \centering
  \includegraphics[width=.6\textwidth]{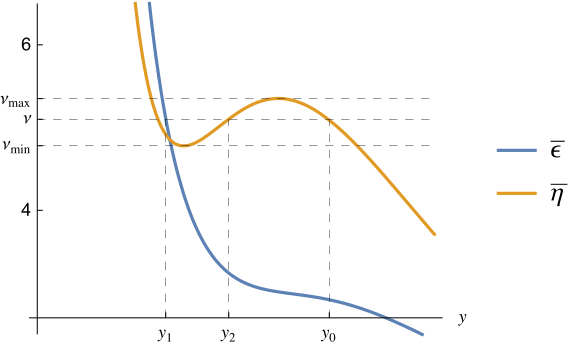}
  \caption{\textit{Shapes of $\bar\epsilon$ and $\bar\eta$ for $0.6021\leq\beta<0.7789$.}}
  \label{fig:shape-2}
\end{figure}

For $\beta$ in this region, $\bar\epsilon$ is still a monotonically decreasing function for
$y\in(0,\infty)$ while $\bar\eta$ is not any longer. The shape of these two slow-roll parameters
are shown in Figure~\ref{fig:shape-2}.
From this graph one can easily read off, given that $\nu_\mathrm{min}<\nu<\nu_\mathrm{max}$, the two slow-roll
regions, one of which is $(y_0,\infty)$ which is trivial, and the other is $(y_1,y_2)$ which
has finite length and is thus the kind of slow-roll region we are searching for. From this
graph we can also see that there are an upper bound and a lower bound for $\nu$ beyond which
there is still only one trivial slow-roll region. In fact, this is a common feature, which
will be justified in Sections~\ref{sec:shape-3},~\ref{sec:shape-4}, and~\ref{sec:shape-5}.
Therefore, all these bounds of $\nu$ corresponding to different $\beta$ render a window
opening to non-trivial slow-roll regions, which shall be plotted in Section~\ref{sec:window}.

\subsubsection{$9/32\leq\beta<0.6021$}
\label{sec:shape-3}

\begin{figure}[!htp]
  \centering
  \includegraphics[width=.6\textwidth]{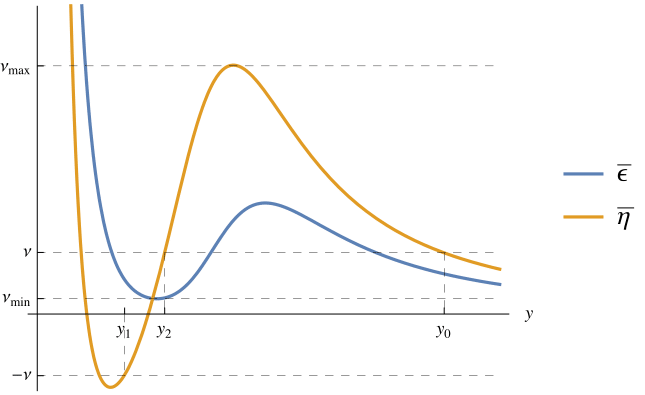}
  \caption{\textit{Shapes of $\bar\epsilon$ and $\bar\eta$ for $9/32\leq\beta<0.6021$.}}
  \label{fig:shape-3}
\end{figure}

For $\beta<0.6021$, both $\bar\epsilon$ and $\bar\eta$ are not monotone functions of $y$ in
$(0,\infty)$, thus we should expect, on the whole, a wider range of $\nu$ that opens to
non-trivial slow-roll regions. In particular, for $\beta\in[9/32,0.6021)$, the potential $v$
is still a monotonically increasing function for $y>0$, which means there is no local minimum
of $v$ in $y\in(0,\infty)$ (the cases that $v$ has a local minimum in $y\in(0,\infty)$ shall
be dealt with in the following two subsections). The typical shapes of $\bar\epsilon$ and
$\bar\eta$ are presented in Figure~\ref{fig:shape-3}. From this figure, we can see that given
$\nu\in(\nu_\mathrm{min},\nu_\mathrm{max})$, there is a non-trivial slow-roll region, $(y_1,y_2)$, apart
from the trivial one, $(y_0,\infty)$.

\subsubsection{$1/4<\beta<9/32$}
\label{sec:shape-4}

\begin{figure}[!htp]
  \centering
  \includegraphics[width=.42\textwidth]{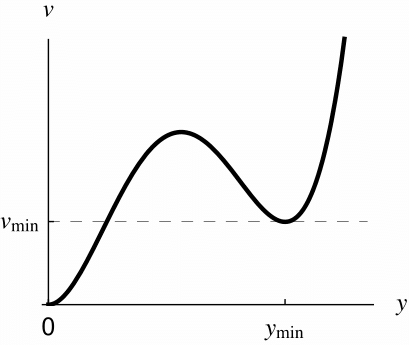} \quad
  \includegraphics[width=.53\textwidth]{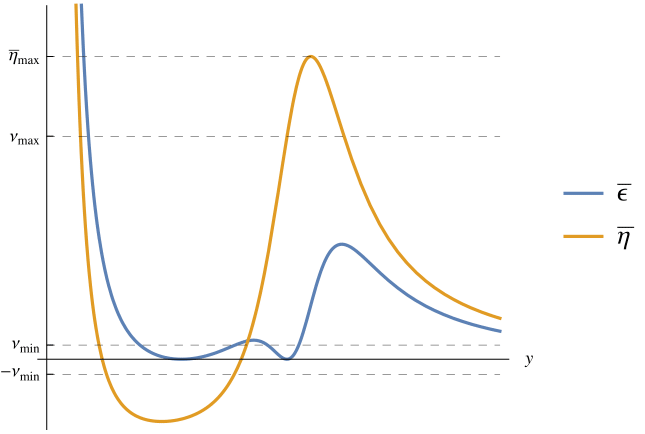}
  \caption[\textit{Graphs illustrating the local minimum and shapes of $\bar\epsilon$ and $\bar\eta$ for $1/4<\beta<9/32$. }]{\textit{The first graph shows the potential $v$ has a local minimum at $y_\mathrm{min}>0$.
  The second illustrates typical shapes of $\bar\epsilon$ and $\bar\eta$ for $1/4<\beta<9/32$.}}
  \label{fig:shape-4}
\end{figure}

In this interval, $v>0$ still holds, but $v'$ is not positive definite any longer. So
$\frac{\partial\bar\epsilon}{\partial y}$ has two more roots which are the roots of $v'$.
When $\beta<9/32$, the potential $v(y)$ has a minimum on the right half $y-v$ plane. In
particular, for $\beta>1/4$ this minimum is a local minimum (see the first graph of
Figure~\ref{fig:shape-4}).
Physically, we prefer the Universe to not being inflating at the minimum of the potential;
the Universe should be reheating and the field should be oscillating.
To ensure this, we look for $|\bar\eta(y_\mathrm{min})|>\nu$, where $y_\mathrm{min}$ is the local
minimum point of $v$. 
Because of this extra filter, the upper bound of $\nu$
(\textit{viz.}, $\nu_\mathrm{max}$) is not necessarily equal to the local maximum of $\bar\eta$ (\textit{viz.},
$\bar\eta_\mathrm{max}$), which is illustrated in the second graph of Figure~\ref{fig:shape-4}.
In that graph, the lower bound of $\nu$, namely $\nu_\mathrm{min}$, is also not at the minimum
of $\bar\epsilon$ which is 0. This is the consequence of the restriction $\bar\eta>-\nu$.

\subsubsection{$0<\beta\leq 1/4$}
\label{sec:shape-5}

\begin{figure}[!htp]
  \centering
  \includegraphics[width=.38\textwidth]{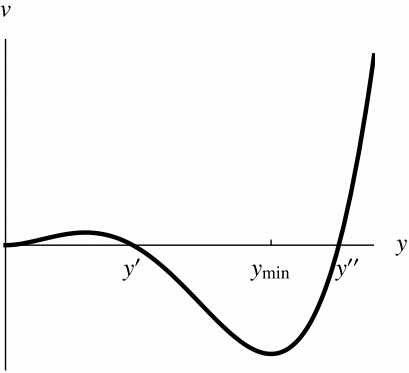} \quad
  \includegraphics[width=.57\textwidth]{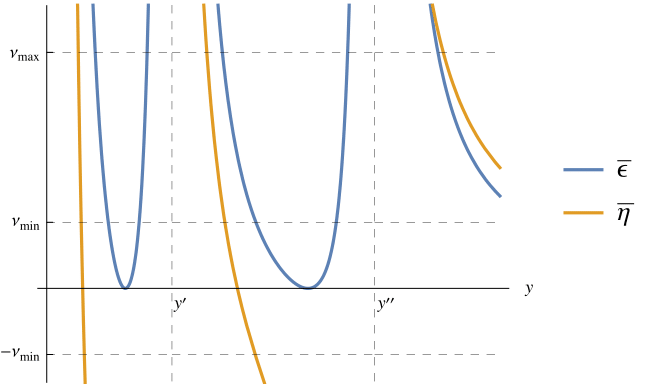}
  \caption[\textit{Graphs illustrating the local minimum and shapes of $\bar\epsilon$ and $\bar\eta$ for $0<\beta<1/4$ }]{\textit{The first graph shows the potential $v$ has a global minimum at $y_\mathrm{min}$.
  The second illustrates typical shapes of $\bar\epsilon$ and $\bar\eta$ for $0<\beta<1/4$.
  The $\beta=1/4$ case is a special case in which $y'$ and $y''$ coincide}.}
  \label{fig:shape-5}
\end{figure}

Finally, when $\beta<1/4$, $v$ in the denominator of~\eqref{eq:d-eps-eta}. This contributes two extra
singularities.
The potential $v$ has a true vacuum in the right half $y-v$ plane, which
is shown in the first graph of Figure~\ref{fig:shape-5}.
At this true vacuum, the potential is negative (or zero for $\beta=1/4$) and thus potential
$v(y)$ has two (or one when $\beta=1/4$) roots, dubbed $y'$ and $y''$ respectively.

As a result, the slow-roll parameters $\bar\epsilon$ and $\bar\eta$ are singular at these
two roots of $v$, which can be seen from the second graph of Figure~\ref{fig:shape-5}.
The bounds of $\nu$, namely $\nu_\mathrm{min}$ and $\nu_\mathrm{max}$ are also denoted in that graph.

\subsection{The window}
\label{sec:window}

\begin{figure}[!htp]
  \centering
  \includegraphics[width=.7\textwidth]{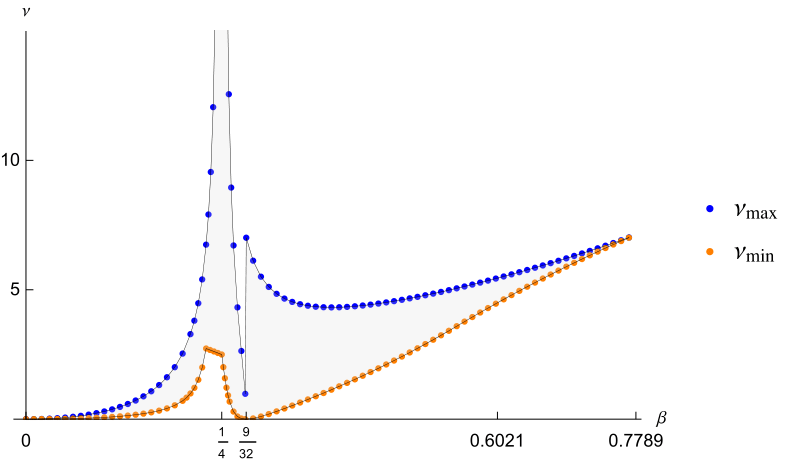}
  \caption{\textit{The window that opens to non-trivial slow-roll potentials.}}
  \label{fig:window}
\end{figure}

Now that we have worked out all possible combinations of $\beta$ and $\nu$ that opens
a window to non-trivial slow-roll potentials whose procedure can be algorithmized by computer
programs, we plot the numerical results in Figure~\ref{fig:window}.
There are two salient features in this figure. First, at $\beta=1/4$, the upper bound
of $\nu$ blows up, which means for any large enough $\nu$ the potential always has a
non-trivial slow-roll region. Second, at $\beta=9/32$, there is a discontinuity (jump)
in the upper bound of $\nu$. This is because, from $\beta>9/32$ to $\beta<9/32$, a local
minimum $y_\mathrm{min}$ appears suddenly in the potential, and the subsequent introduction of
the extra constraint $|\bar\eta(y_\mathrm{min})|>\nu$ which ensures that the Universe does not
inflate at the minimum of the potential makes the bound for $\nu$ discontinuous.

\subsubsection{The likelihood of potentials being slow-roll}
\label{sec:likelihood}

Since we have made the assumption that the energy scale of inflation is at $M_{GUT}$, from
\eqref{eq:define-nu} we see that $\nu$ is around $10^{-8} a^{-2}$. If $a$ is of order $\mathcal{O}(1)$,
$\nu$ should be an extremely small number, then Figure~\ref{fig:window} indicates the likelihood of
$(\beta, \nu)$ falling into the window must be quite small. To provide some evidence of our rough
estimation, we made a statistical experiment using Monte Carlo method. We first assume that $a$
and $b$ are normally distributed and exponentially distributed variables respectively,
\begin{equation}
  a \sim \mathcal{N}(0, 1), \quad \textrm{and} \quad b \sim \textrm{Exp}(1).
\end{equation}
We than randomize $(a, b)$ over $400 000$ samples and found $312$ slow-roll instances,
which evaluates to a likelihood of approximately $0.078$\%.

A second method of estimating the success rate is to compute the ratio of the area within the window to the area of the parameter space being surveyed.
Because the window goes off to infinity, we introduce a cutoff, corresponding to excluding the region around $a=0$.
For a cutoff at $\nu = 10000$, the ratio is $\sim0.15$\%;
for a cutoff at $20000$, the ratio is $\sim0.14$\%;
for a cutoff at $40000$, it is $\sim0.13$\%.
This ratio is similar to the result obtained using the Monte Carlo method.

\section{Multi-field models}
\label{sec:multi-field}
In this section, we investigate which potentials accommodate the slow-roll conditions for inflation with two fields.
The form of the potentials we have is
\begin{eqnarray}
	\nonumber
	V(\tilde{\varphi},\tilde{\chi}) &=& \frac{\mu^2}{2} M^2 \tilde{\varphi}^2 +  \frac{\rho^2}{2} M^2 \tilde{\chi}^2 +
	a_1 M \tilde{\varphi}^3 + a_2 M \tilde{\varphi}^2 \tilde{\chi} + a_3 M \tilde{\varphi} \tilde{\chi}^2 + a_4 M \tilde{\chi}^3 \\
	&& + b_1 \tilde{\varphi}^4 + b_2 \tilde{\varphi}^2 \tilde{\chi}^2 + b_3 \tilde{\chi}^4~,
	\label{eq:multi-potential}
\end{eqnarray}
where the masses $\mu M$ and $\rho M$ for the fields $\tilde\phi$ and $\tilde\chi$ are defined in terms of $M$, the GUT mass, the $a_i M$ are cubic couplings, and the $b_i$ are quartic couplings.
We will assume that the masses are around the GUT scale ($\sim 10^{16}$ GeV).
We motivate the quartic potential from a Wilsonian perspective wherein higher order terms are suppressed by the energy scale at which new physics enters.
We assume this is the string scale or Planck scale ($\sim 10^{19}$ GeV).
Terms higher than quartic order, as they are suppressed by this higher energy scale, are neglected in the analysis.
The coefficients $a_i$ and $b_i$ are order one numbers.
The terms that appear in~\eqref{eq:multi-potential} are dictated by the fact that we demand all slow-roll potentials to be bounded from below.
With this in mind, the $a_i$ can be positive or negative and the $b_i$ are positive.
When both fields tend to $-\infty$, the quartic terms should have no odd powers in any of the two variables.
We can rescale~\eqref{eq:multi-potential} similarly to what we did in the single field case.
With $\tilde{\varphi} = M \varphi$ and $\tilde{\chi} = M \chi$, we have
\begin{eqnarray}
	\nonumber
	v(\varphi,\chi) &=& \frac{\mu^2}{2} \varphi^2 +  \frac{\rho^2}{2} \chi^2 +
	a_1 \varphi^3 + a_2 \varphi^2 \chi + a_3 \varphi \chi^2 + a_4 \chi^3 \\
	&& +  b_1 \varphi^4 + b_2 \varphi^2 \chi^2 + b_3 \chi^4.
	\label{eq:multi-dimensionless}
\end{eqnarray}
Now all parameters and variables in potential $v$ are dimensionless and it is sensible
to talk about the magnitude of parameters. Note that $V = M^4 v$.
These methods can readily be generalized to having more scalar fields.
We simply require that the superpotential is renormalizable and bounded from below.
As adding more scalars and studying the potentials explicitly in the finite field case is computationally more intensive, we do not extend the analysis beyond the two field level in this work.
When searching for minima in the potential one will encounter both false and true vacua.
We allow for slow-roll regions around false vacua (local minima) and not only the true vacua (global minima). The reason that we allow for both
false and true vacua for slow-roll in two field case is based on computational and physical grounds. If we only look for global minima to test our
slow-roll constraints, we firstly need to find global minima using methods such as steepest gradient descent subject to some arbitrary
initial conditions. This will greatly increase the computational time and render the numerical test impossible in a reasonable time frame.
Moreover, on physical grounds as long as the false vacuum is sufficiently long lived, the Universe may be in a metastable state.
We do not analyze the lifetime to exclude short lived false vacua as this analysis depends on details of, say, particle physics and the presence of other nearby minima.

\subsection{Slow-roll conditions for multi-field inflation}
\label{subsec:multi-condition}
It is important to discuss the slow-roll conditions for multi-field inflation models as they are fundamentally different from those of single field case. The conditions are discussed in detail in~\cite{Cosmo_Chap43}. We shall demand the following:
\begin{align}
	\nonumber	
	\epsilon &\equiv -\frac{\dot{H}}{H^2} = 3\left(\frac{\dot{\phi_i}^2}{V}\right)
	=\frac{M_\mathrm{Pl}^2 (\partial_i V)^2}{2V^2} \ll 1~,\\
	\xi &\equiv \sqrt{\hat{V_1}\cdot\overleftrightarrow{V_2}\cdot
	\overleftrightarrow{V_2}\cdot\hat{V_1}} \ll 1~,
	\label{def:multi-SR-conditions}
\end{align}
with
\begin{eqnarray}
	\hat{V_1} \equiv \frac{\partial_i V}{|\partial_i V|}~,
	\overleftrightarrow{V_2} \equiv	\frac{M_\mathrm{Pl}^2(\partial_i\partial_jV)}{V}~,
	\label{def:V1-V2}
\end{eqnarray}
for fields $\vphi = (\varphi,\chi)$.
Here the conditions are derived from the approximation $3H\dot{\phi}_i \approx
-\partial_i V$, which is essentially the consistent second slow-roll condition.
This comes down to neglecting $\ddot{\phi}_i$ compared to
$\partial_iV$. But when comparing two vectors, it is sensible only to compare their
norms. Therefore we have the \textit{strong second slow-roll condition} $|\ddot{\phi}_i| \ll |\partial_iV|$. The reason it is called the strong second slow-roll condition is because
its smallness implies
\begin{equation}
	\eta \equiv \hat{V_1} \cdot \overleftrightarrow{V_2} \cdot \hat{V_1} \ll 1~,
	\label{def:multi-eta}
\end{equation}
where $\eta$ is defined to be
\begin{equation}
	\frac{1}{\epsilon H}\frac{d\epsilon}{dt} = 4\epsilon -
	2 \hat{V_1} \cdot \overleftrightarrow{V_2} \cdot \hat{V_1}
	= 4\epsilon - 2\eta \ll 1~.
\end{equation}
Therefore the slow gradient flow by the fields defined in~\eqref{def:multi-SR-conditions}
is not the only way to get a slowly-varying quasi-de Sitter expanding phase.

\subsection{Numerical tests}
\label{subsec:num-test}
In this section, we numerically determine whether a potential of
the form~\eqref{def:V1-V2} satisfies the slow gradient flow condition in~\eqref{def:multi-SR-conditions}.
Because we now have the free parameters $\vec{a}$ and $\vec{b}$, we will adopt the Monte Carlo paradigm to characterize the shapes of potentials and quantify the rate of success.

\subsubsection{Setup for numerics}
\label{subsubsec:num-setup}

The experiment is set up as follows.
\begin{enumerate}
\item\label{sec:dist_choice}
The coefficients $\vec{a}$ and $\vec{b}$ in cubic and quartic terms in~\eqref{eq:multi-dimensionless}
are first sampled from a uniform distribution within range $[-3,3]$ and $[0,5]$ respectively.
In addition, we also sampled the $\vec{a}$ coefficients from a Gaussian distribution with mean $0$
and variance $1$, and $\vec{b}$ coefficients from a exponential distribution with $\lambda = 1$.

Let us briefly justify these choices of parameters.
The experiments with the uniform distribution are performed
in the spirit of Monte Carlo simulations, where parameters are chosen essentially at random. 
The choice of the uniform distribution is further justified by the fact that we do not know the region where
slow-roll solutions reside in the seven dimensional parameter space of the potential coefficients.
On the other hand, the choice of normal distribution with particular mean and variance will center our
data around that mean and therefore may miss possible slow-roll regions.
The choice of uniform distribution reflects the fact that we have no knowledge on the region of slow-roll
samples within the parameter space \emph{a priori}. In addition, the parameters are chosen to be of $O(1)$ with respect
to GUT scale. This comes from the fact that the higher order terms of the potential do not get corrections from
quantum gravity effects, thereby, the potential is written in this particular quartic form.
Note that this polynomial potential allows vertices that
mix the two inflatons. This rules out the models such as assisted inflation where
potential takes steep exponential~\cite{Cosmo_Chap44} due to the fact that our potentials are polynomials.

In the second set of experiments,
the Gaussian distribution for $\vec{a}$ is motivated by the Central Limit Theorem.
If we suppose our coefficients can be observed, the averages of $n$ measurements
of each coefficient then approach Gaussian distribution when $n \to \infty$. Meanwhile, 
the mean and variance are chosen on the grounds of naturalness.
\textit{A priori}, the coefficients should be order one numbers at the scale determined by the
masses of the inflatons, which we set to GUT scale.
Therefore, our choice of Gaussian distribution for $\vec{a}$ can be seen from previous arguments.
On the other hand, we have $\vec{b}$ follow a Gamma distribution. We demand each of the elements
of $\vec{b}$ to be positive  in order to ensure that the potential is bounded from below. 
Just as the Gaussian distribution is a maximum entropy probability distribution for positive and
negative real numbers, the Gamma distribution is  the maximum entropy probability
distribution for positive real numbers. The Gamma distribution therefore becomes the natural candidate for
selecting coefficients. In particular, we use an exponential distribution with $\lambda = 1$, which
is a Gamma distribution with shape parameter $k = 1$ and scale parameter $\theta = \lambda$.
Again, the choice of $1$ is motivated on the grounds of naturalness.
The Central Limit Theorem requires large $n$.
It is not clear that this applies when only a small number of scalar fields participate in inflation, so the comparison between the two possibilities is useful.

\item For each set of random coeffcients $\vec{a}$ and $\vec{b}$, we search for a point
that satisfies conditions~\eqref{def:multi-SR-conditions} within a particular range for
fields $\phi_i$ by using the {\sc Mathematica}~\cite{mathematica} function\footnote{Of course,
we wrote more code than just this one line function.}
\begin{equation}
  \textbf{FindInstance[$<${\it slow-roll conditions}$>$, $\{\varphi$, $\chi\}$]}.
  \label{eq:findInst}
\end{equation}
The search region within field space is rectangular with origin in the middle.
The size of both sides of this region is twice the maximum of distances between origin and
any stationary points of the potential. \footnote{To be precise, we used \texttt{GroebnerBasis[]} to solve
for zeros of the gradients of the potentials to find the extrema. However, this method actually turns the cubic polynomials
into higher power polynomials (sometimes as high as 9-th order) thus making numerical solution highly sensitive to small change of
gradient. For future work, we suggest to use \texttt{NSolve[]} directly instead and this might change the results slightly.}
This is justified because we want a slow-roll region
that is near a stationary point and the potential becomes steep far out from the origin in our
potentials that are bounded from below.
We do not want to falsely classify solutions as slow-roll simply by virtue of the fact the denominator, which is determined by the value of the potential, is large near infinity in field space.

\item For particular conditions in~\eqref{def:multi-SR-conditions}, we have observational
constraints of
\begin{equation}
  \epsilon < 0.01 \quad \textrm{and} \quad \xi < 0.01
  \label{eq:SR-observational}
\end{equation}
from measurements of scalar spectral
index $n_s$ that directly restricts slow-roll parameters. We also note that since the $\xi$
condition implies $\eta$, the results from imposing $\xi$ should be smaller than those from
$\eta$.

\item The inflaton mass parameters are defined as $\mu = m_{\varphi}/M_\mathrm{GUT}$ and
$\rho = m_{\psi}/M_\mathrm{GUT}$, where $m_{\varphi}$ and
$m_{\psi}$ are the inflaton masses. Here, we set them both to be of GUT scale, so $\mu$ and
$\rho$ are of order $O(1)$. 
With the mass parameters fixed, we take $N=314000$ uniformly distributed random coefficient samples,
$(\vec{a}_1, \vec{b}_1), \dots, (\vec{a}_N, \vec{b}_N)$. To be precise, we first of all generate 157000
samples using the {\sc Mathematica} function {\bf RandomReal[\{-3, 3\}, 4]} for $\vec{a}$
and {\bf RandomReal[\{0, 5\}, 3]} for $\vec{b}$. Then we apply the slow-roll conditions
in~(\ref{def:multi-SR-conditions}) to these coefficients using the approach and constraint described
in~(\ref{eq:findInst}) and (\ref{eq:SR-observational}) to obtain instances\footnote{The {\bf FindInstance}
function in {\sc Mathematica} is not capable of finding all desired slow-roll instances due to the mathematical complications of the slow-roll
  conditions and the internal algorithms designed for this task in {\sc Mathematica}. Our experimental comparison of {\bf FindInstance} with the more comprehensive but slower {\bf NSolve} function indicate that results from using the two options are similar.}
 of relative minima that satisfy the necessary conditions.
By noticing that the potential~\eqref{eq:multi-dimensionless} is symmetric under
the following transformation for $\mu=\rho$,
\begin{equation}
  \varphi \leftrightarrow \chi ,\quad a_1 \leftrightarrow a_4 ,\quad a_2 \leftrightarrow a_3
  ,\quad b_1 \leftrightarrow b_3,
\end{equation}
we apply this symmetry to the slow-roll coefficients found previously in the $157 000$ samples.
That is equivalently getting all slow-roll coefficients in $314 000$ random samples. In addition, this
is also done for $145 000$, or equivalently, $290 000$ samples drawn from Gaussian/Gamma distributions with means and shape parameters chosen as unity.

\end{enumerate}

\subsubsection{Numerical results}
\label{subsubsec:num-results}
\begin{figure}
  \centering
  \includegraphics[width=.8\textwidth]{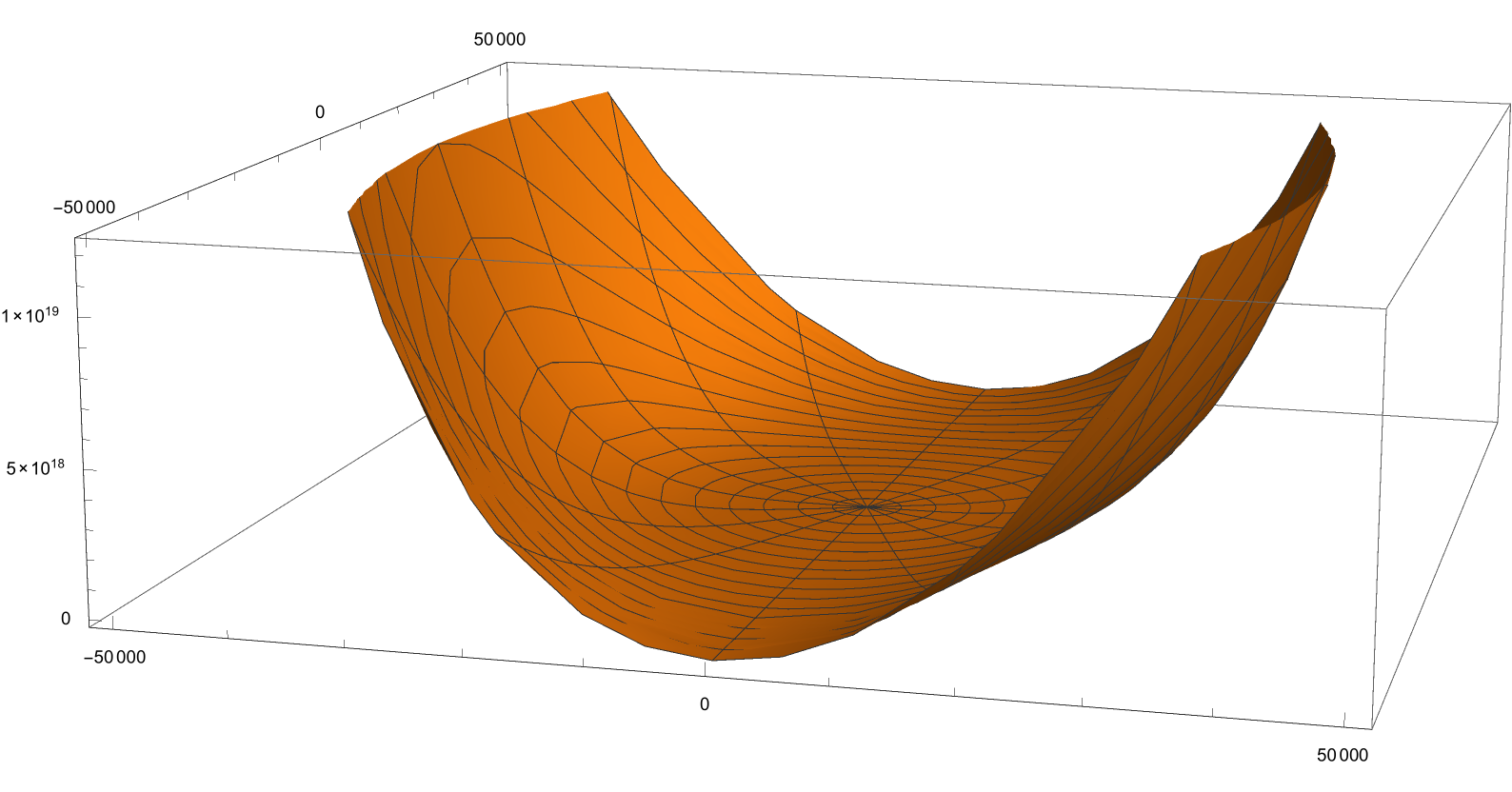}
  \includegraphics[width=.8\textwidth]{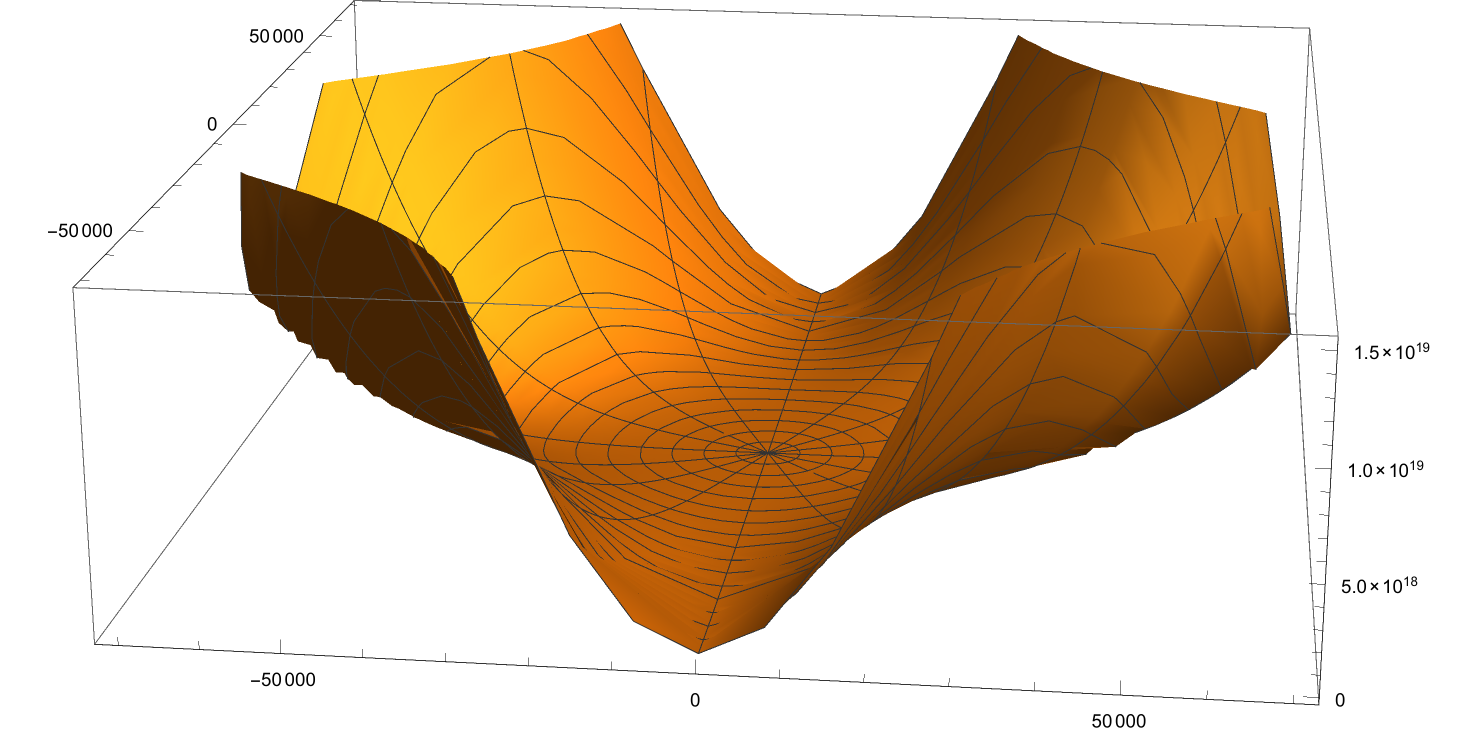}
  \caption[\textit{Two of the 76 slow-roll potentials for the uniform distribution in Table~\ref{tab:sl-samples}}]{\textit{Two of the 76 slow-roll potentials for the uniform distribution in Table~\ref{tab:sl-samples}.
  They have coefficients $\bm{a}=(1.85634, -2.75233, 0.59967, 0.655031)$, $\bm{b}=(4.61147, 3.281, 0.00123865)$
  and $\bm{a}=(0.675461, -0.286123, 2.6534, 1.6393)$, $\bm{b}=(0.294909, 4.25653, 0.00658533)$ respectively.
  The point instances found in their slow-roll regions are at $(-46745.5, -426)$ and $(-64074., -193.667)$,
  and the potential values there are around $2.2\times 10^{19}$ and $5\times 10^{18}$. However, the farthest
  stationary points are at $(-0.091192, -396.138)$ and $(-0.311574, -186.663)$.}}
  \label{fig:typical-v}
\end{figure}

Of all the slow-roll potentials we obtained from choosing coefficients using both the uniform and Gaussian distributions, all of the
  slow-roll points found with function {\bf FindInstance} have their distances from the origin greater
  than $30 000$. The absolute values of potentials corresponding to these point instances have order of
  magnitude above ${\mathcal O}(10^{16}$). Figure~\ref{fig:typical-v} depicts two typical slow-roll
potentials we have found for the uniform distribution.

\begin{eqnarray}
	\nonumber
  v(\varphi,\chi) &=& \frac{\varphi^2}{2} + \frac{\chi^2}{2}
  + 1.85634 \varphi^3 - 2.75233 \varphi^2\chi + 0.59967 \varphi\chi^2 + 0.655031 \chi^3 \\
  && + 4.61147 \varphi^4 + 3.281 \varphi^2\chi^2 + 0.00123865 \chi^4,
	\label{eq:typical-v-1}
\end{eqnarray}
has three real stationary points,
\begin{equation}
  (-0.091192, -396.138), \quad (-0.028419, -0.497133) \quad\textrm{and}\quad (0, 0),
\end{equation}
the furthest of which is around 400 units of distance away from the origin. However, the slow-roll
point we found is at $(-46745.5, -426)$, which is far beyond the region we expected. The potential
at the slow-roll point is $2.2\times 10^{19}$, and the gradient is $-1.9\times 10^{15}$ and $-6.1\times 10^{12}$
in $\varphi$ and $\chi$ directions respectively. One can check that, the two slow-roll parameters,
\begin{equation}
  \epsilon = 0.0037 \quad\textrm{and}\quad \xi = 0.0055,
\end{equation}
satisfy the constraints in~(\ref{eq:SR-observational}). However, from the careful inspection of this specific
example, we can see that the slow-roll (and its adjacent region) satisfy the $\epsilon \ll 1$ and $\eta \ll 1$ conditions because the potential at this distance from the origin is large and the two slow-roll parameters are both roughly
inversely proportional to some power of the potential. All the other slow-roll instances we found in our random
samples are similar to this specific example. We stated in Section~\ref{sec:single} that we wanted to exclude those
regions where~(\ref{eq:slow-roll-condition}) was satisfied simply due to the largeness of the potential.
However, now that these are the only slow-roll regions we have found for the two-field case, the best statistics
we can get is from this kind of slow-roll potential.

\begin{figure}[!htp]
  \centering
  \includegraphics[width=.8\textwidth]{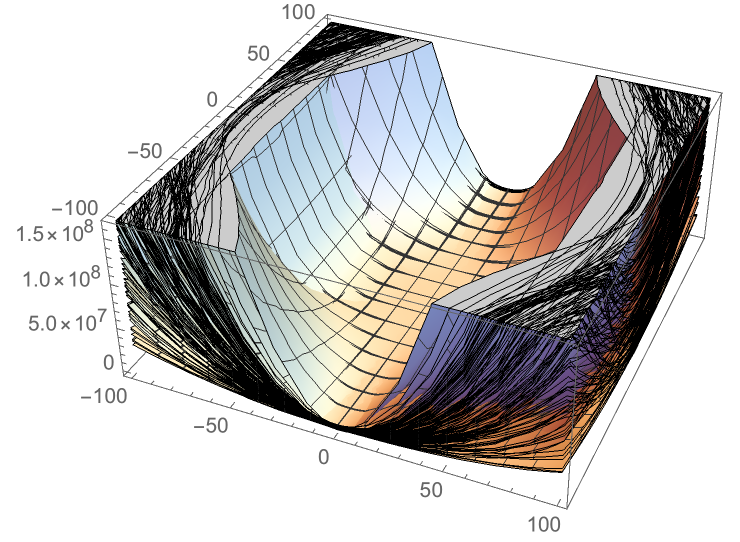}
  \includegraphics[width=.8\textwidth]{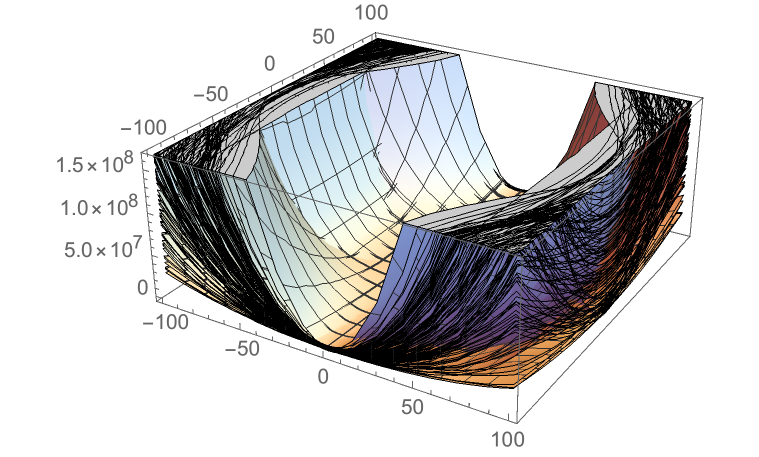}
  \caption[\textit{Stack of all found potentials for both uniform and Gaussian distributions}]{\textit{The top figure gives the stack of all found potentials for uniform distribution and the bottom one gives that for Gaussian distribution.}}
  \label{fig:v-stack}
\end{figure}

To get a better idea about the slow-roll potential, we stack all the successful potentials in a single plot
in Figure~\ref{fig:v-stack}. As the figure shows, all the slow-roll potentials have a steep and a flat direction. 
Although multi-field models typically allow for the presence of isocurvature modes, Figure~\ref{fig:v-stack} indicates that such modes
are heavily suppressed as the two-field inflation appears to be dominated by a single flat direction.
This qualitative shape of the potentials recalls those found in models of hybrid inflation type~\cite{Cosmo_Intro23}.
Generic initial conditions in the neighborhood of a minimum would first fix the field in the direction perpendicular to the valley (the steep direction), and then the field would roll along the valley to the minimum at the bottom.
Unlike the potential for hybrid inflation discussed in the introduction, informed by our choices of random coefficients, the parameters that appear in the potentials here are all typically of order one with respect to the GUT scale.

The statistics of slow-roll potentials we found in potentials from the uniform and Gaussian/Gamma
distribution are listed in Table~\ref{tab:sl-samples}.
From Table~\ref{tab:sl-samples}, we can see that in all randomly generated coefficients, around 0.05\% of them correspond
to slow-roll potentials and 0.1\% for Gaussian/Gamma distribution. This supplies a \textit{lower bound}
for the percentage of relative minima that accommodate the slow-roll conditions for inflation with two scalar fields. 
\begin{table}
\centering
\begin{tabular}{|c|c|c|c|c|c|}
  \hline
  Distribution & Range of $\bm{a}$ & Range of $\bm{b}$ & Samples & Slow-Roll & Percentage \\
  \hline
  Uniform & $[-3,3]$ & $[0,5]$ & 157,000 & 76 & 0.05\% \\
  \hline
  Gaussian Gamma & $[-\infty, \infty]$ & $[0, \infty]$ & 145,000 & 131 &0.1\% \\
  \hline
\end{tabular}
\caption{\textit{Slow-roll potentials found in random sample potentials of chosen distributions}} \label{tab:sl-samples}
\end{table}

\begin{figure}[!htp]
\begin{subfigure}{.5\textwidth}
\centering
\includegraphics[width=.8\textwidth]{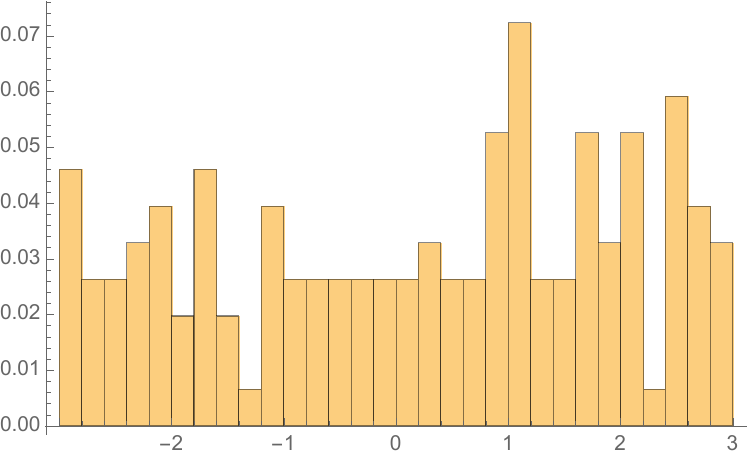}
\caption{\textit{$a_1$ \& $a_4$}}
\end{subfigure}
\begin{subfigure}{.5\textwidth}
\centering
\includegraphics[width=.8\textwidth]{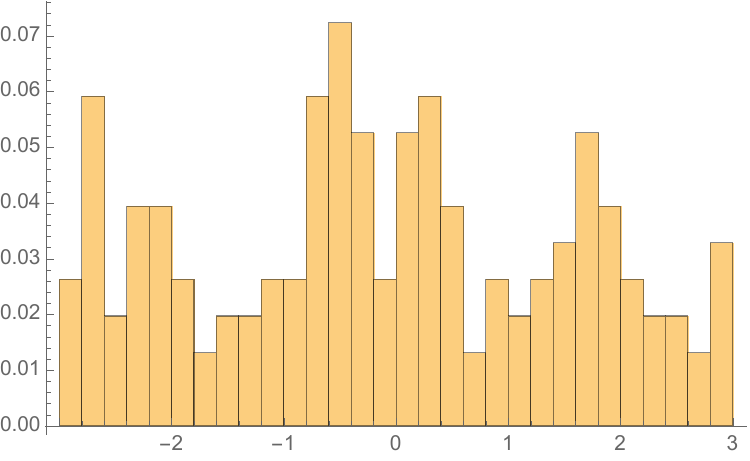}
\caption{\textit{$a_2$ \& $a_3$}}
\end{subfigure}
\begin{subfigure}{.5\textwidth}
\centering
\includegraphics[width=.8\textwidth]{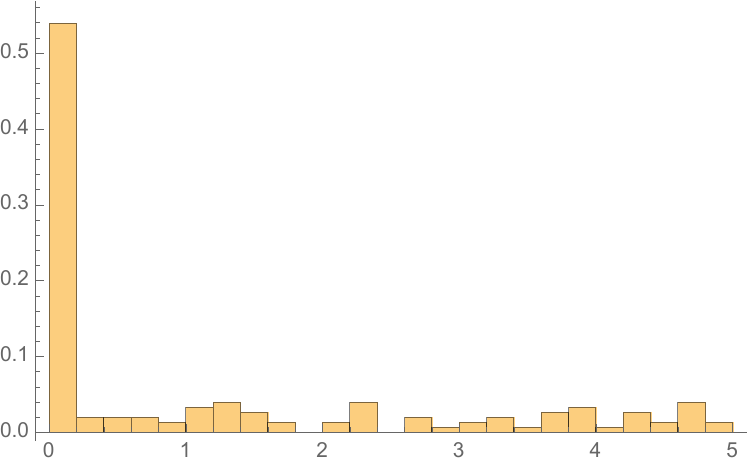}
\caption{\textit{$b_1$ \& $b_3$}}
\end{subfigure}
\begin{subfigure}{.5\textwidth}
\centering
\includegraphics[width=.8\textwidth]{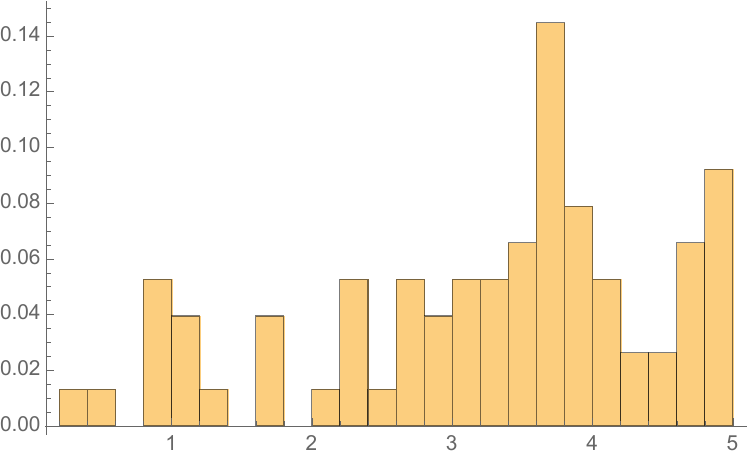}
\caption{\textit{$b_2$}}
\end{subfigure}
\caption[\textit{Distribution histogram for $\vec{a}$ and $\vec{b}$ coefficients drawn from uniform distribution.}]{\textit{Distribution histogram for $\vec{a}$ and $\vec{b}$ coefficients drawn from uniform distribution.
The $x$-axes are values for $\vec{a}$ and $\vec{b}$ and the $y$-axes are for probabilities for those values to occur.}}
\label{fig:uniform}
\end{figure}

We also draw distributions of values of coefficients $(\vec{a}, \vec{b})$ that correspond to slow-roll potentials
in Figure~\ref{fig:uniform}.
These figures show that in spite of the uniform distribution that we presume as priors for $a_i$ and $b_i$,
the slow-roll conditions pick the coefficients according to the mass distributions in these histograms.
The distributions of all components of $\vec{a}$ and $\vec{b}$ deviate noticeably from the uniform distribution.
This indicates that the slow-roll conditions set some constraints on the values of coefficients in a
probabilistic sense. Deviations may also be a product of having a low number of slow-roll regions, only around $70$, despite starting with a minimum of $150 000$ samples. A greater computational analysis would be required to obtain more concrete statistics for the distribution of slow-roll regions.
It is notable that the results are roughly the same irrespective of what distribution we employ to select the coefficients.

\begin{figure}[!htp]
\begin{subfigure}{.5\textwidth}
\centering
\includegraphics[width=.8\textwidth]{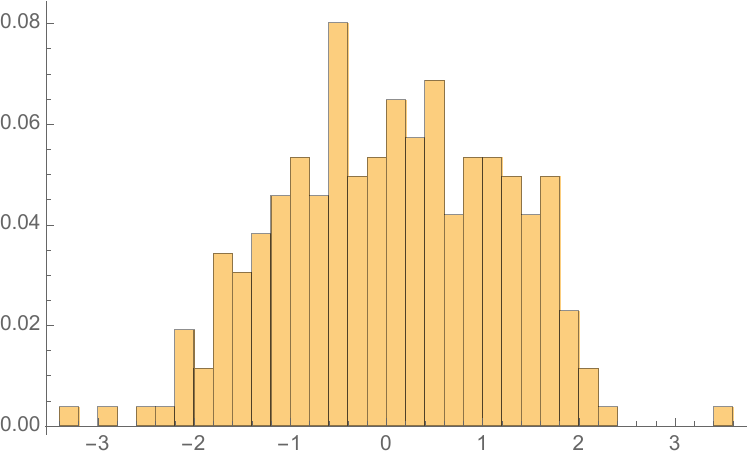}
\caption{\textit{$a_1$ \& $a_4$}}
\end{subfigure}
\begin{subfigure}{.5\textwidth}
\centering
\includegraphics[width=.8\textwidth]{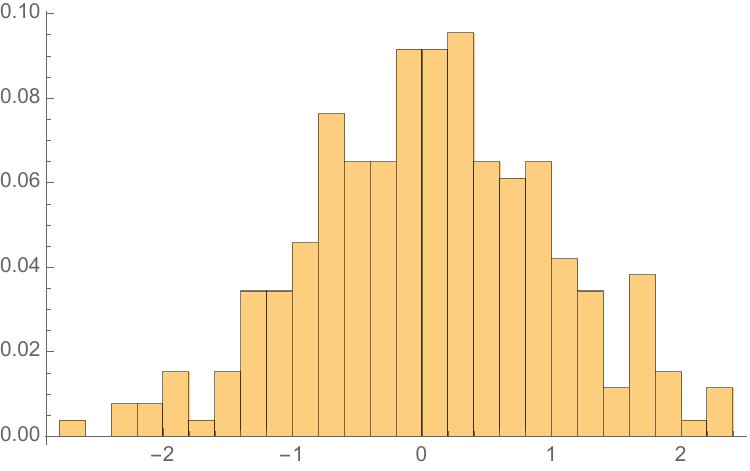}
\caption{\textit{$a_2$ \& $a_3$}}
\end{subfigure}
\begin{subfigure}{.5\textwidth}
\centering
\includegraphics[width=.8\textwidth]{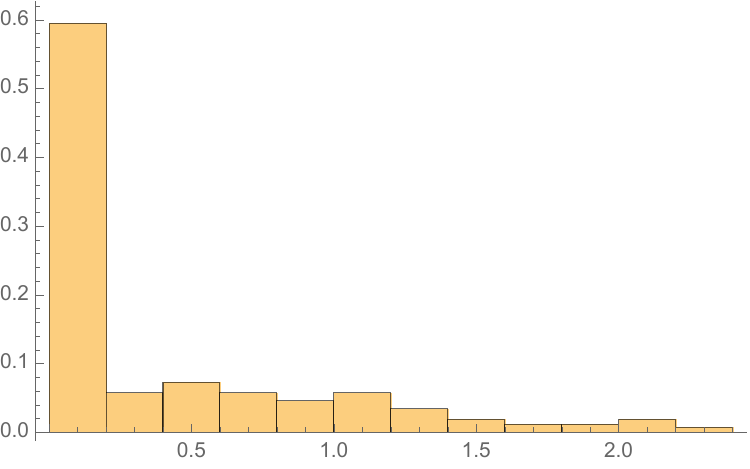}
\caption{\textit{$b_1$ \& $b_3$}}
\end{subfigure}
\begin{subfigure}{.5\textwidth}
\centering
\includegraphics[width=.8\textwidth]{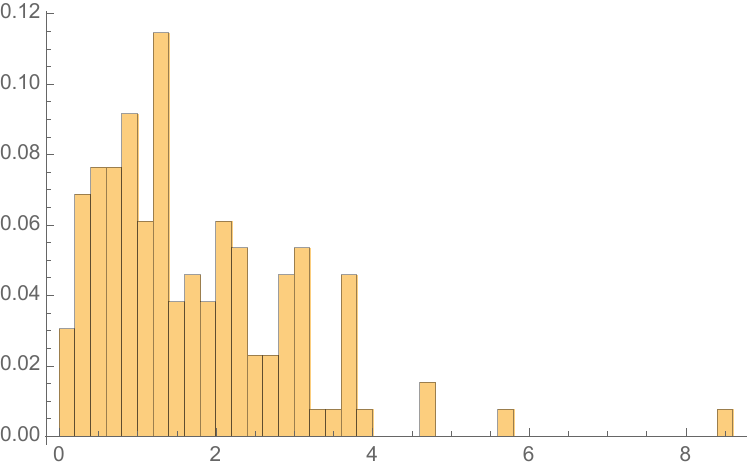}
\caption{\textit{$b_2$}}
\end{subfigure}
\caption[\textit{Distribution histogram for $\vec{a}$ and $\vec{b}$ coefficients drawn from Gaussian/Gamma distribution.}]{\textit{Distribution histogram for $\vec{a}$ and $\vec{b}$ coefficients drawn from Gaussian/Gamma distribution.
The $x$-axes are values for $\vec{a}$ and $\vec{b}$ and the $y$-axes are for probabilities for those values to occur.}}
\label{fig:gauss}
\end{figure}

In addition to the results from uniform distribution, we also present the histogram plots for coefficients
of slow-roll potentials for Gaussian/Gamma distribution in Figure \ref{fig:gauss}. We can see from the plot
that the Gaussian nature of the plot still is still present. This shows that the slow-roll conditions we choose
respects the Gaussianity of initial samples.

Preliminary experiments with varying the mass parameters for the scalar fields over several orders of magnitude do not significantly change the percentages of slow-roll solutions.

\begin{figure}[!htp]
	\begin{subfigure}{.5\textwidth}
		\centering
		\includegraphics[width=.8\textwidth]{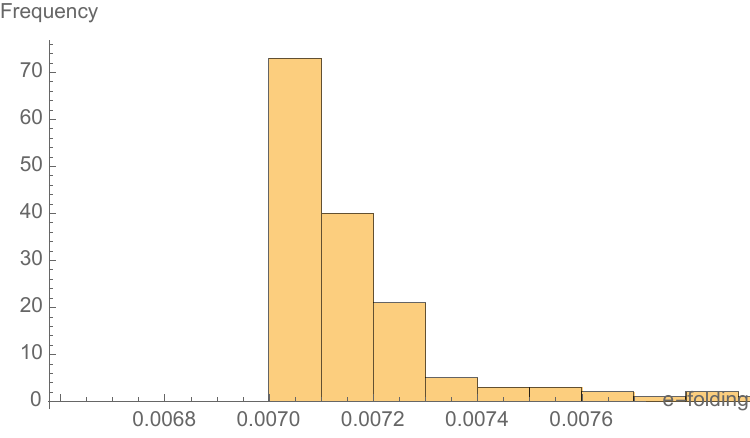}
		\caption{\textit{The e-foldings for the top potential in Fig. \ref{fig:typical-v}.}}
	\end{subfigure}
	\begin{subfigure}{.5\textwidth}
		\centering
		\includegraphics[width=.8\textwidth]{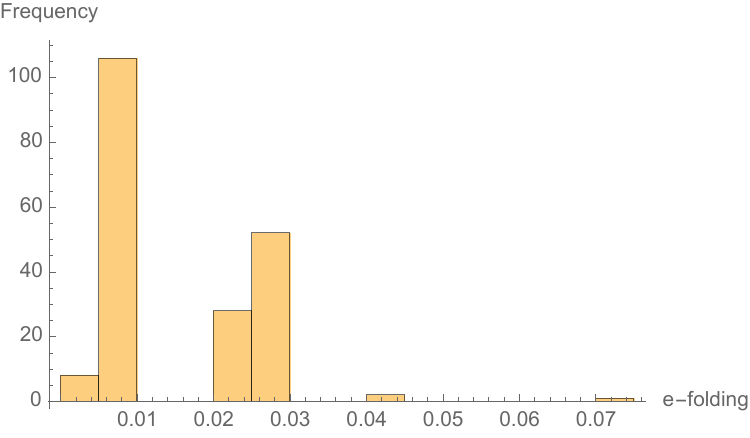}
		\caption{\textit{The e-foldings for the bottom potential in Fig. \ref{fig:typical-v}.}}
	\end{subfigure}
\caption{\textit{Histograms corresponding to different initial conditions
	for sample potentials in Figure \ref{fig:typical-v}.} }
\label{fig:efold}
\end{figure}

Finally, in Figure~\ref{fig:efold} we present two histograms for $e$-foldings from different initial
conditions for the two potentials in Figure~\ref{fig:typical-v}.

The calculation for $e$-foldings is set up as follows.
\begin{itemize}
	\item The number of $e$-foldings is a function of the initial conditions. In a single field model, we may compute
	\begin{equation}
	N(\phi) = \frac{1}{M_\mathrm{Pl}^2} \int_{\phi_*}^{\phi_i} d\phi\ \frac{V(\phi)}{V'(\phi)} \simeq \int_{\phi_*}^{\phi_i} \frac{d\phi}{\phi}\, \frac{1}{\eta} ~,
	\end{equation}
	where $\phi_i$ is the initial value and $\phi_*$ is the critical value corresponding to the minimum of the potential.
	\item Since all the potentials in the two field case that we consider have a steep direction and a flat direction, we set component of the field
	vector in the steep direction to a fixed value and  perform the integration along the flat direction to
	simplify this into an effective single field problem.
	\item Let us denote the steep direction $\phi$ and the flat direction by $\psi$. For each potential, we find
	its extrema and identify $(\phi_*,\psi_*)$ 
	 with the field configuration of the global minimum. Then we want to see how the perturbation in the path
	 of integration affect the values of $e$-foldings. This is done by sampling through the final values $\phi_f$,
	 for the integration in the range of $\phi_f \in [\phi_*-|100\phi_*|,\phi_*+|100\phi_*|]$ in steps of $|\phi_*|$. The field that is being integrated over is then  the component in the flat direction $\psi$.
	 The starting value $\psi_i$ is half the value of the distance between the nearest extremum to the global minimum.
	 For example, the bottom potential in Figure~\ref{fig:typical-v} has its global minimum at $(-186.663,-0.311574)$,
	 and the nearest extremum in the flat direction is at $(-0.152773,-0.0523367)$. We then choose the initial and final
	 value for the integration in the flat direction to be $(-186.663,-90)$.
\end{itemize}
With the previous setup, we obtain the histograms in Figure~\ref{fig:efold}, and we see that the values of $e$-foldings
are a few magnitudes smaller compared to the phenomenological required value of $\sim 60$. Therefore, even if
we find certain potentials supports slow-roll at particular points, the potential can fail to sustain period of inflation
long enough to have the experimental values of $e$-folding. It is important to note that these are preliminary results.
The focus of our paper is the preliminary issue of whether slow-roll is even possible given a random potential. 

\section{Discussion and outlook}\label{sec:discussion}

With the advent of Big Data in theoretical physics and ever-increasing computational power, we are gaining further glimpses into the various landscapes of theories ranging from string vacua to cosmological scenarios.
In this paper, we have been motivated by the question of the probability of having slow-roll inflation within the landscape of effective potentials for inflatons.

We started with the case of a single field with the most generic form of the potential up to degree four, subject to the constraints of slow-roll. Here, there are only two  parameters, which we have dubbed $\beta$ and $\nu$, and which can be expressed in terms of the couplings.
We can solve the problem numerically to arrive at Figure~\ref{fig:window}.
The figure depicts a non-trivial region of parameter values which satisfy the slow-roll conditions.

With two inflatons, the situation is understandably more intricate.
Here, up to degree four, there are seven parameters.
The slow-roll conditions then translate to a polynomial system in the fields and in the parameters.
This is then a problem in a potential landscape sculpted by these parameters.
We find that the slow-roll conditions for multi-field are insensitive to the distribution we have used, \textit{i.e.}, we find that they give the same percentage of slow-roll instances.
Moreover, initial experiments that change the mass scale yield similar results.
The inflatons for slow-roll inflation traverse far into field space so that the slow-roll conditions are satisfied in part due to the largeness of $V$.
The potential has a characteristic shape in which the fields would roll first down a steep direction and then follow a valley to the minimum.
Based on this shape, the generation of isocurvature modes is not expected.

In the single field case, the condition $V'\, V''' \ll V^2\, M_\mathrm{P}^{-4}$ accounts for constraints on the running of the spectral index.
It would be reasonable to develop the equivalent third derivative condition for
multi-field case and include this within the framework of random potentials.
Preliminary analysis of example potentials that admit slow-roll minima suggest that an insufficient number of $e$-foldings transpires.
While we can exclude certain minima as being unable to support the necessary number of $e$-foldings, the suitability of other minima for this purpose depends critically on the choice of initial conditions for the inflaton fields.
In this work we do not perform a systematic study of the number of $e$-foldings
for each of the potentials that support slow-roll inflation.
There is no obvious \textit{a priori} selection criteria for this informed by realistic string constructions of the Standard Model. 
We defer a more detailed investigation for future work. 

In general, with an arbitrary number of inflatons and a potential up to a specified degree, the slow-roll constraints will produce a large polynomial system with still larger number of parameters. 
For example, in the heterotic string Standard Models to which we alluded in the introduction, the contribution to the number of moduli fields come from the geometry --- roughly the sum of the Hodge numbers --- and from the bundle --- roughly the number of endomorphisms~\cite{Cosmo_Chap45}. For the $(3,3)$ Calabi--Yau threefold studied in~\cite{Cosmo_Chap3, CY_Chap24} the total number of moduli is $6+19 = 25$.

Because it is doubly exponential in the number of parameters, the usual Groebner basis~\cite{Cosmo_Chap46} approach to analyzing such systems will soon become rather prohibitive, and even numerical algebraic geometry~\cite{Cosmo_Chap39} will find this challenge daunting. Our approach of randomization over parameters is thus the standard technique, and the statistics over the landscape is an enlightening overview of how special or generic our universe is. Aided by work done in \cite{Cosmo_Chap41,Cosmo_Chap42}, one could find a relationship between the general statistics of successful inflation in random polynomial potentials with the general statistics of stationary points in the same potentials. They note that the variance of positive real roots both increases with the degree of the polynomial and converges with sample size. Knowing this may give, for example, a convenient saturation limit of test samples, shifting a greater proportion of computational resource to just the search for relationships between minima and slow roll regions. Finding this would allow us to derive properties of the inflationary landscape from the statistics of the minima themselves. More work is needed hover in developing actual ``search and test" techniques for inflation within these minima.

Large-$N$ field models are difficult to study when looking for successful inflation. Although the upper bound probability of critical points of a random polynomial grows as $\sim \frac{N}{2}$ \cite{p1}, the probability of inflation in the large $N$ field cases appears to decrease as seen in work done in \cite{Cosmo_Chap27}, which provides a heuristic argument that in a multifield landscape probability distribution to realize slow-roll inflation looks like 
\begin{equation}
P(N_e) = AN^{-\alpha}_e ~,
\end{equation}
where $A$ is exponentially small and $\alpha\sim 3$. (See~\cite{Cosmo_Intro22, p4, p5, p6}.)
Since out work was limited to two and three fields, it is difficult to make conclusive statements or extrapolate the probabilities we have obtained to the large $N$ limit.
We expect that the more fields we have, the harder it becomes to realize slow-roll conditions using a random potential.

It has recently been conjectured that constraints on cosmological inflation are two-fold~\cite{vafa1,vafa2}:
(1) the distance traversed in field space during inflation should be ${\cal O}(1)$; and
(2) the gradient of the potential during inflation satisfies
\begin{equation}
\frac{\nabla V}{V} = \frac{\sqrt{G^{ij} \partial_i V \partial_j V}}{V} \gtrsim {\cal O}(1) ~, \qquad \partial_i = \frac{\partial}{\partial\phi^i} ~.
\end{equation}
In the single field case, whereas we can choose parameters to satisfy (1), generically, we have not found examples that satisfy (2) within our allowed window from Figure~\ref{fig:window}.
In the two field case, in the examples we have examined, the shape of the potential fails to satisfy the second criterion by roughly two orders of magnitude.

One should mention also that there is a branch of mathematics known as random algebraic geometry where the usual quantities such topological invariants and cohomology, which govern the physics, have their analogues in the stochastic sense. It will certainly be interesting to study polynomial systems arising from the potential landscape under this light. Furthermore, with the advancement of machine learning techniques, it may be instructive to see how a neural network could be applied to finding slow-roll points in a potential. Rather than teaching a neural network to look for successful points of slow-roll inflation, one could look at the methodology these neural networks use to minimize the cost function. For a single feature in a neural net, one would use a simple gradient descent method to recursively find a local minima. This is equivalent to looking for the local minima of a single field polynomial potential. Typically, however, a neural network has an enormous amount of features, or input variables, which all determine the cost functions, and thus minimizing this cost function, is equivalent to finding a critical point of a multivariate polynomial, or in our language, a multifield potential. Neural networks do this rather efficiently, and thus one possible avenue is to tweak how these networks find critical points and ask them to find critical points that allow for slow-roll inflation.
This is work in progress.

\section*{Acknowledgements}
We thank Cyril Matti for collaboration during an early stage of this work.
YHH would like to thank the Science and Technology Facilities Council, UK, for grant ST/J00037X/1, the Chinese Ministry of Education, for a Chang-Jiang Chair Professorship at NanKai University as well as the City of Tian-Jin for a Qian-Ren Scholarship, and Merton College, Oxford, for her enduring support.
VJ and LP are supported by the South African Research Chairs Initiative, which is funded by the Department of Science and Technology and the National Research Foundation of South Africa.
YX is support by the Doctoral Studentship of City University of London.
DZ is supported by the China National Natural Science Foundation under contract No.~11105138 and 11575177.
DZ is also indebted to Chinese Scholarship Council (CSC) for support.

\end{document}